\documentstyle[aps,prl]{revtex}
\begin{document}
\preprint{}
\draft
\title{A  Novel Superstring in Four Dimensions and Grand Unification}
\author{B. B. Deo}
\address{Physics Department, Utkal
University, Bhubaneswar-751004, India.} 
\maketitle
\begin{abstract}
A string in four dimensions is constructed by supplementing it with  
forty four Majorana fermions. The central charge is 26.
The fermions are grouped in
such a way that the resulting action is supersymmetric.
The energy momentum and current generators satisfy the super-Virasoro
algebra.  The tachyonic ground state 
decouples from the physical states. GSO projections are necessary
for proving modular invariance. Space-time supersymmetry provides
reasons to discard the tachyons and is substantiated for modes of zero
mass. The symmetry group of the  model descends to the low energy
standard model group $SU (3) \times SU_L (2) \times U_Y (1)$ through
the Pati-Salam group. Left right symmetry is broken spontaneously and 
the mass of the tau neutrino is calculated to be about 1/25 electron volt.

\end{abstract}

\section{Introduction}
String theory was invented \cite{bd1} as a sequel to dual
resonance models \cite{bd2} to explain the properties of 
strongly interacting particles in four dimensions. Assuming
 a background gravitational field and
demanding Weyl invariance, the Einstein equations of general
relativity could be deduced. It was believed that about these
classical solutions one can expand and find the quantum
corrections. But difficulties arose at the quantum level.
Eventhough the strong interaction amplitude obeyed crossing, 
it was no longer unitary. There were anomalies and ghosts. 
Due to these compelling reasons it was necessary for the open 
string to live in 26
dimensions \cite{bd3,bd10}. At present the most successful theory is a ten 
dimensional superstring on a Calabi-Yau manifold or an orbifold.
However, in order to realise the programme of the string unification of all
the four types of interactions, one must eventually arrive at a
theory in four flat space-time dimensions, with N=1 supersymmetry
and chiral matter fields. This paper is an attempt in that direction.

A lot of research has been done to construct four 
dimensional strings  \cite{bd99}, specially in the 
latter half of the eighties. Antoniadis
et al\cite{ia} have constructed a four dimensional superstring supplemented 
by eighteen real fermions in trilinear coupling. The central charge of the
construction is 15. Chang and Kumar \cite{dc} have discussed 
Thirring fermions, but again with the central charge at 15. Kawai et al
\cite{hk} have also considered four dimensional models in a different context
than the model proposed here. None of these models makes contact with the 
standard  model.

In section II, we give the details of the supersymmetric model.
Section III gives the usual quantization and super-Virasoro algebra is
deduced in the section IV. Bosonic states are constructed in Section
V. Fadeev- Popov ghosts are introduced and the
BRST charge is explicitly given in section VI. Ramond states 
have been worked out in section VII. In section VIII,   
the mass spectrum of the model and the necessary GSO projections to
eliminate the half integral spin states are introduced.  In section
IX, we show that these projections are necessary to prove the modular 
 invariance of the model. Space-time supersymmetry algebra is
 satisfied and is shown to exist
for the zero mass modes in section X. In section XI we show how the 
chain $SO(44)\to SO(11)\to SO(6)\times SO(5) \to SU_C (3)\times SU_L (2)
\times U_Y(1)$ is possible in this model. We
calculate that the Pati-Salam group $SU(4) \times SU_L (2) \times SU_R (2)$ 
breaks at an intermediate mass $M_R \simeq 5 \times 10^{14}$ GeV giving the left-handed
neutrino a small mass, which has now been observed in the top sector.

The literature on string theory is very vast and exist in most
text books on the subject. The references serve only as a guide
to elucidate the model.

\section{The Model} 
The model essentially consists of 26 vector bosons of an open (closed)
string in which there are
the four bosonic coordinates of four dimensions and there are 
fortyfour Majorana fermions representing the remaining 22 
bosonic coordinates \cite{bd11}. We divide them into four 
groups .
They are labelled by $\mu=0,1,2,3$ and
each group contains 11 fermions. These 11 fermions are again
divided into two groups, one containing six and the other five. 
For convenience, in one group we have $j=1,2,3,4,5,6$,
and in the other,
$k=1,2,3,4,5$.

The string action is
\begin{equation}
S=-\frac{1}{2 \pi} \int d^2 \sigma \left [ {\partial}_{\alpha} 
X^{\mu}{\partial}^{\alpha} X_{\mu}-i \bar{\psi}^{\mu,j}
\rho^{\alpha} {\partial}_{\alpha}\psi_{\mu,j}-i\phi^{\mu,k} 
\rho^{\alpha}{\partial}_{\alpha}\phi_{\mu,k}\right ]\;,
\end{equation}
$\rho^{\alpha}$ are the two dimensional Dirac matrices
\begin{equation}
\rho^0=\left( \begin{array}{cc} 0 & -i\\
i & 0 \end{array} \right)\;,\;\;\;\;\;
\rho^1=\left( \begin{array}{cc} 0 & i\\
i & 0 \end{array} \right)
\end{equation}
and obey 
\begin{equation}
\{\rho^{\alpha}, \rho^{\beta}\}=-2 \eta_{\alpha \beta}\;.
\end{equation}
String co-ordinates $X^{\mu}$ are scalars in $(\sigma ,\tau)$ space and
vectors in target space. Similarly $\psi^{\mu ,j}$ are spinors in
$(\sigma ,\tau)$ space and vectors in target space. 

In general we follow the notations and conventions of reference
\cite{bd5} whenever omitted by us. $X^{\mu}(\sigma,\tau)$ are 
the string coordinates. The Majorana fermions $\psi$'s and
$\phi$'s are decomposed in the basis
\begin{equation}
\psi=\left( \begin{array}{c} \psi_- \\ \psi_+ \end{array}
\right)\;,
\;\;\;\mbox{and}\;\;\;\; 
\phi=\left( \begin{array}{c}\phi_- \\ \phi_+ \end{array}
\right)\;.
\end{equation}
The nonvanishing commutation and anticommutations are
\begin{equation}
[\dot{X}^{\mu} (\sigma,\tau),X^{\nu}(\sigma^{\prime},\tau)]
=-i \pi \;\delta(\sigma-\sigma^{\prime})\eta^{\mu \nu}
\end{equation}
\begin{equation}
\{\psi_A^{\mu}(\sigma,\tau), \psi_B^{\nu}(\sigma^{\prime},\tau)\}
=\pi\; \eta^{\mu \nu}\; \delta_{AB}\; \delta(\sigma-\sigma^{\prime})
\end{equation}
\begin{equation}
\{\phi_A^{\mu}(\sigma,\tau), \phi_B^{\nu}(\sigma^{\prime}, \tau)\}
=\pi\; \eta^{\mu \nu}\; \delta_{AB}\; \delta(\sigma-\sigma^{\prime})
\end{equation}

The action is invariant under infinitesimal transformations
\begin{equation}
\delta X^{\mu}=\bar \epsilon\; \left ( \sum_j \psi^{\mu,\;j}
+i\sum_k\phi^{\mu,\;k}\right )
\end{equation}
\begin{equation}
\delta\psi^{\mu,\;j}=-i \rho^{\alpha}\; \partial_{\alpha} X^{\mu}
\;\epsilon
\end{equation}
\begin{equation}
\delta \phi^{\mu,\;k}=+\rho^{\alpha}\;\partial_{\alpha}\; X^{\mu}
\;\epsilon
\end{equation}
where $\epsilon$ is an infinitesimally constant anticommuting
Majorana spinor. The commutator of the two supersymmetry
transformations gives a spatial translation, namely
\begin{equation}
[\delta_1,\delta_2]X^{\mu}=a^{\alpha}\;{\partial}_{\alpha}X^{\mu}
\end{equation}
and
\begin{equation}
[\delta_1,\delta_2]\Psi^{\mu}=a^{\alpha}{\partial}_{\alpha}
\Psi^{\mu}
\end{equation}
where 
\begin{equation}
a^{\alpha}=2i\;\bar \epsilon_1\; \rho^{\alpha}\; \epsilon_2
\end{equation}
and
\begin{equation}
\Psi^{\mu}=\sum_j\psi^{\mu,\;j}+i\sum_k\phi^{\mu,\;k}
\end{equation}
In deriving this, the Dirac equation for the spinors have been
used. The Noether super-current is
\begin{equation}
J_{\alpha}=\frac{1}{2} \rho^{\beta}\;\rho_{\alpha}
\;\Psi^{\mu}\; \partial_{\beta} X_{\mu}
\end{equation}

We now follow the standard procedure. The light cone components
of the current and energy momentum tensors are
\begin{equation}
J_+=\partial_+X_{\mu}\; \Psi^{\mu}_+
\end{equation}
\begin{equation}
J_-=\partial_-X_{\mu}\;\Psi^{\mu}_-
\end{equation}
\begin{equation}
T_{++}=\partial_+X^{\mu} \partial_+X_{\mu}+\frac{i}{2}
\psi_+^{\mu,\;j}\;\partial_+\psi_{+\mu,\;j}+\frac{i}{2}
\phi_+^{\mu,\;k}\;\partial_+\phi_{+\mu,\;k}
\end{equation}

\begin{equation}
T_{--}=\partial_-X^{\mu} \partial_-X_{\mu}+\frac{i}{2}
\psi_-^{\mu,\;j}\partial_-\psi_{-\mu,\;j}+\frac{i}{2}
\phi_-^{\mu,\;k}\partial_-\phi_{-\mu,\;k}
\end{equation} 
where $\partial_{\pm}=\frac {1}{2} (\partial_{\tau}\pm 
\partial_{\sigma})$.

To proceed further we note that in equation (8) and (14) we could have 
taken $-i$ instead of $+i$. We now introduce a phase factor
$\eta_{\phi}$ to replace $`i'$ in these equation. $\eta_{\phi}$
depends on the number $n_{\phi}$ of $\phi_{s}$ (or its quanta), in 
a given individual term. Explicitly $\eta_{\phi} = (-1)^{1/4
  n_{\phi}(n_{\phi} - 1) + \frac{1}{2}}$. $\eta_{\phi} = i$ if
$n_{\phi} = 1$ reproducing $`i'$ in the above equations. But
$\eta_{\phi}^{2} = -1$ if $n_{\phi} = 0$ where two $\phi$'s have
been contracted away and $\eta_{\phi}^{2} = 1$ if $n_{\phi} = 2$.

One now readily calculates the algebra
\begin{eqnarray}
\{ J_{+} (\sigma), J_{+}(\sigma^{\prime})\} && = \pi
\delta(\sigma - \sigma^{\prime}) T_{++} (\sigma)  \nonumber \\
\{ J_{-} (\sigma), J_{-}(\sigma^{\prime})\} && = \pi
\delta(\sigma - \sigma^{\prime}) T_{--} (\sigma) \nonumber \\
\{ J_{+} (\sigma), J_{-}(\sigma^{\prime})\} && = 0 
\end{eqnarray}

The time like components of $X^{\mu}$ are eliminated by the use of
Virasoro constraints $T_{++} = T_{--} = 0$. In view of equation
(20), we postulate that
\begin{equation}
0 = J_{+} = J_{-} = T_{++} = T_{--}
\end{equation}

$J_{+}$ is a sum of a real and imaginary term, The real term is
a sum of six mutually independent $\psi^{\mu, j}$ 's and the
imaginary term, the five mutually independent $\phi^{\mu, k}$ 's. It
will be shown in Section V, that $J_{+} = 0$ constraint excludes all
the eleven time like components of $\psi$'s and $\phi$'s from the
physical space.

\section{Quantization}

As usual the theory is quantized  ($\alpha_o^{\mu} = p^{\mu}$), with
\begin{equation}
X^{\mu}=x^{\mu}+p^{\mu}\tau +i\sum_{n\neq 0}\frac{1}{n}\alpha^{\mu}_n
\exp^{-i n\tau} cos(n\sigma), \nonumber
\end{equation}
or
\begin{equation}
\partial_{\pm}X^{\mu}=\frac{1}{2}\sum_{-\infty}^{+\infty}
\alpha_n^{\mu}\; e^{-in(\tau\pm\sigma)}
\end{equation}
\begin{equation}
[\alpha_m^{\mu},\alpha_n^{\nu}]=m\; \delta_{m+n}\;
\eta^{\mu \nu}
\end{equation}

While discussing the mass spectrum, it will be more illuminating to
consider the closed string. The related additional quantas here and in
wherever occurs will be denoted by attaching a tilde. For instance
\begin{equation}
\partial_{-}X^R_{\mu}=\sum_{-\infty}^{+\infty}\alpha^{\mu}_n
e^{-2in(\sigma-\tau)}
\end{equation}
\begin{equation}
\partial_{+}X^L_{\mu}=\sum_{-\infty}^{+\infty}\tilde{\alpha}^{\mu}_n
e^{-2in(\sigma+\tau)}
\end{equation}
The transition formulas for closed strings can be easily effected. 
We consider the open string. 
We first choose the Neveu-Schwarz (NS) \cite{bd4}
boundary condition. Then the mode expansions of the fermions are
\begin{equation}
\psi_{\pm}^{\mu,j}(\sigma,\tau)=\frac{1}{\sqrt 2}
\sum_{r\in Z+\frac{1}{2}}b_r^{\mu,\;j}e^{-ir (\tau\pm\sigma)}
\end{equation}

\begin{equation}
\phi_{\pm}^{\mu,k}(\sigma,\tau)=\frac{1}{\sqrt 2}
\sum_{r\in Z+\frac{1}{2}}b_r^{\prime\,\mu,\;k}e^{-ir (\tau \pm \sigma)}
\end{equation}
\begin{equation}
\Psi_{\pm}^{\mu,j}(\sigma,\tau)=\frac{1}{\sqrt 2}
\sum_{r\in Z+\frac{1}{2}}B_re^{-ir (\tau\pm\sigma)}
\end{equation}
The sum is over all the half-integer modes.
\begin{equation}
\{b_r^{\mu,j}, b_s^{\nu, j^{\prime}}\}=\eta^{\mu \nu}\;\delta_
{j,j^{\prime}}\; \delta_{r+s}
\end{equation}
\begin{equation}
\{b_r^{\prime\,\mu,k}, b_s^{\prime\,\nu, k^{\prime}}\}=\eta^{\mu \nu}\;\delta_
{k,k^{\prime}}\;\delta_{r+s}
\end{equation}
\begin{equation}
\{B_r^{\mu}, B_s^{\nu}\}=\eta^{\mu \nu}\;
\delta_{r+s}\;.
\end{equation}

\section{Virasoro Algebra}

Virasoro generators \cite{bd6} are given by the modes of the
energy momentum tensor $T_{++}$ and Noether current $J_+$,
\begin{equation}
L_m^M=\frac{1}{\pi}\int_{-\pi}^{+\pi} d\sigma\; e^{im\sigma}\; T_{++}
\end{equation}
\begin{equation}
G_r^M=\frac{\sqrt 2}{\pi}\int_{-\pi}^{+\pi} d\sigma\; e^{ir\sigma}\; J_{+}
\end{equation}
`$ M $' stands for matter. In terms of creation and annihilation
operators
\begin{equation}
L_m^M=L_m^{(\alpha)}+L_m^{(b)}+L_m^{(b')}
\end{equation}
where
\begin{equation}
L_m^{(\alpha)}=\frac{1}{2}\sum_{n=-\infty}^{\infty}:
\alpha_{-n}\cdot\alpha_{m+n}:
\end{equation}

\begin{equation}
L_m^{(b)}=\frac{1}{2}\sum_{r=-\infty}^{\infty}
(r+\frac{1}{2}m) :b_{-r}\cdot b_{m+r}:
\end{equation}

\begin{equation}
L_m^{(b')}=\frac{1}{2}\sum_{r=-\infty}^{\infty}
(r+\frac{1}{2}m):b'_{-r}\cdot b'_{m+r}:
\end{equation}

In each case normal ordering is required. The single dot
implies the sum over all qualifying indices. 
The current generator is
\begin{equation}
G^M_r= \sum_{n = - \infty}^{\infty} \alpha_{-n} \cdot (b_{r+n} +
\eta_{\phi} b^{\prime}_{r+n}) = \sum_{n=-\infty}^{\infty} \alpha_{-n}\cdot
(b_{r+n}+i\;b'_{r+n})=\sum_{n=\infty} \alpha_n\cdot B_{r+n}
\end{equation}
Following from eqn. (33) the Virasoro algebra is
\begin{equation}
[L_m^M, L_n^M]=(m-n) L_{m+n}^M+A(m)\;\delta_{m+n}
\end{equation}
Using the relations
\begin{equation}
\left [L^M_m,\alpha_n^{\mu} \right ]=-n\alpha^{\mu}_{n+m}
\end{equation}
\begin{equation}
\left [L^M_m,B_n^{\mu} \right ]=-(n+\frac{m}{2})B^{\mu}_{n+m}
\end{equation}
we get, also
\begin{equation}
[L_m^M, G^M_r]=\left (\frac{1}{2}m-r\right ) G_{m+r}^M
\end{equation}
The anticommutator $\{G^M_r, G^M_s\}$ is obtained directly or 
by the use of the Jacobi identity
\begin{equation}
[\{G^M_r,G^M_s\},L_m^M]+\{[L_m^M,G^M_r],G^M_s\}+\{
[L_m^M,G^M_s],G^M_r\}=0
\end{equation}
which implies, consistent with equations (34) and (35),
\begin{equation}
\{G^M_r,G^M_s\}=2 L_{r+s}^M+B(r)\delta_{r+s}
\end{equation}
$A(m)$ and $B(r) $ are normal ordering anomalies. Taking the 
vacuum expectation value in the Fock ground state $|0,0\rangle $ 
with four momentum $ p^{\mu}=0$ of the commutator $[L_1,L_{-1}]$
and $[L_2, L_{-2}]$, it is easily found that
\begin{equation}
A(m)=\frac{26}{12}(m^3-m) = \frac{C}{12}(m^{3}-m)
\end{equation}
and using the Jacobi identity
\begin{mathletters}
\begin{equation}
B(r)=\frac{A(2r)}{2r}
\end{equation}
\begin{equation}
B(r) = \frac{26}{3}\left (r^2-\frac{1}{4}\right ) = \frac{C}{3} \left (
  r^2 - \frac{1}{4} \right )
\end{equation}
\end{mathletters}
The central charge $C=26$. This is what is expected.
Each bosonic coordinate contribute 1 and each fermionic
ones contribute $1/2$, so that the total central charge is +26.

For closed strings there will be another set of tilded 
generators satisfying the same algebra.

\section{Bosonic States}

A physical bosonic state $\Phi$ which should be invariant under $SO(6)\times SO(5)$ 
internal symmetry group and can be convinently constructed by operating the
generators $L$'s and $G$'s on the vacuum. They satisfy
\begin{equation}
 L_m^M\; \mid \Phi \rangle=0\;\;\;\;\;\;\;\;\;m>0
\end{equation}
\begin{equation}
 G_r^M\; \mid \Phi \rangle=0\;\;\;\;\;\;\;\;\;r>0
\end{equation}
These conditions enable to exclude the time like quanta from the
physical spectrum. Specialising to a rest frame we
write the conditions (48) as
\begin{equation}
\frac{1}{2} p^{0} \alpha_{m}^{0} \mid \Phi \rangle \; + \; {\rm
  (terms~quadratic~in~osillators)} \mid \Phi \rangle = 0
\end{equation}
In this frame, the physical states are generated effectively by the
space components of the oscillators only; so that
$\alpha_{m}^{0}\mid\Phi \rangle = 0$ following from the constraint
that the energy momentum tensor vanishes. Using the condition (49),
\[ \left [ G_{r}^M, \alpha_{m}^{0} \right ] \mid \Phi \rangle = m
b_{m+r}^{0} \mid \Phi \rangle = 0 \]
means
\begin{equation}
 (b_{r}^{0,1} + \cdot \cdot \cdot + b_{r}^{0,6})\mid \Phi \rangle  = 0
\end{equation}
\begin{equation}
 (b_{r}^{\prime 0,1} + \cdot \cdot \cdot + b_{r}^{0,5})\mid \Phi
 \rangle  = 0
\end{equation}

 $b_{r}^{0,1}$ to $b_{r}^{0,6}$  or $b_{r}^{\prime 0,1}$ to
$b_{r}^{\prime 0,5}$ are all independent anahilation operators for $r
\; > \; 0$ and there is no relation between them. Therefore $\Phi$
decouple from  the eleven time like components
$b_{r}^{j}$'s or $b^{\prime k}_{r}$ `s, for, otherwise the equality to
zero in equations (51) and (52) cannot be achieved. Thus the vanishing 
of the energy-momentym tensor and the current excludes all the time
like components from the physical space. No negative norm state will
show up in the physical spectrum and at the same time preserve $SO(6)
\times SO(5)$ internal symmetry. The above arguments are only qualitative.

Let us make a detailed investigation to ensure that there are no
negative norm physical states. We shall do this by constructing the
zero norm states or the `null' physical states. Due to the  GSO condition,
which we shall study later, the physical states will be obtained by
operation of the product of even number of G's. So the lowest state
above the tachyonic state is
\[ \mid \Psi \rangle = L_{-1} \mid \chi_{1} \rangle + G_{-1/2}
G_{-1/2} \mid \chi_{2} \rangle \]
But $ G_{-1/2} G_{-1/2} = \frac{1}{2} \{  G_{-1/2}, G_{-1/2} \} =
L_{-1}$. Without loss of generality, the state is
\begin{equation}
\mid \Psi \rangle = L_{-1} \mid \tilde{\chi} \rangle
\end{equation}
This state to be physical, it must satisfy $( L_{0} - 1) \mid \Psi
\rangle = 0$ which is true if $L_{0} \mid \tilde{\chi} \rangle = 0$. The
norm $\langle \Psi \mid \Psi \rangle = \langle \tilde{\chi}\mid L_{1} L_{-1} \mid
\tilde{\chi} \rangle = 2 \langle \tilde{\chi} \mid L_{0} \mid \tilde{\chi} \rangle 
= 0$. Let us consider the next higher mass state
\[ \mid \Psi \rangle = L_{-2} \mid \chi_{1} \rangle + L_{-1}^{2} \mid
\chi_{2} \rangle + ( G_{-3/2} G_{-1/2} + \lambda G_{-1/2} G_{-3/2} )
\mid \chi_{3} \rangle + G_{-1/2} G_{-1/2}G_{-1/2}G_{-1/2} \mid
\chi_{4} \rangle  + \cdots \]
It can be shown that $G_{-3/2} G_{-1/2} \mid \tilde{\chi} \rangle = (
\beta_{1} L_{-1}^{2} + \beta_{2} L_{-2}) \mid \tilde{\chi}
\rangle$. The coefficients $\beta_{1}$ and $\beta_{2}$ can be
calculated by evaluating $\left [ L_{1}, G_{-3/2} G_{-1/2} \right ]
\mid \tilde{\chi} \rangle$ and $\left [ L_{2}, G_{-3/2} G_{-1/2} \right ]
\mid \tilde{\chi} \rangle$. $G_{-1/2}^{4}$ is proportional to
$L_{-1}^{2}$. So, in essence, we have the next excited state as
\begin{equation}
\mid \Psi \rangle = \left ( L_{-2} + \gamma L_{-1}^{2} \right ) \mid
\tilde{\chi} \rangle
\end{equation}
The condition $(L_{0} - 1)\mid\Psi\rangle=0$ is satisfied if $(L_{0} + 1) 
\mid \tilde{\chi} \rangle = 0$. Further the physical state condition
$L_{1} \mid \Psi \rangle = 0$ gives the value of $\gamma = 3/2$. The
norm is easily obtained as 
\begin{equation}
\langle \Psi \mid \Psi \rangle = \frac{1}{2} (C - 26)
\end{equation}
This is negative for $C < 26$ and vanishes for $C=26$. So the critical 
cenbtral charge is 26. It is easily checked that $L_{2} \mid \Psi
\rangle$ also vanishes for $C=26$.

To find the role of $b$ and $b^{\prime}$ modes, let us calculate the norm
of the following state with $p^{2} = 2$
\begin{equation}
(L_{-2} + 3/2 L_{-1}^{2} ) \mid 0, p \rangle = \left (
  L_{-2}^{(\alpha)} + \frac{3}{2} L_{-1}^{(\alpha)^{2}} \right ) \mid, 
0, p \rangle + \left (  L_{-2}^{(b)} + \frac{3}{2} L_{-1}^{(b)^{2}}
\right ) \mid 0, p \rangle + \left (  L_{-2}^{(b^{\prime})} +
  \frac{3}{2} L_{-1}^{(b^{\prime})^{2}} \right ) \mid 0, p \rangle
\end{equation}
The norm of the first term is equal to $-11$ as calculated in reference
\cite{bd5}.

Noting that $L_{-1}^{(b)} \mid 0,p\rangle = L_{-1}^{(b^{\prime}}) \mid
  0, p \rangle = 0$;  $L_{-2}^{(b)} = \frac{1}{2} b_{-3/2} \cdot
  b_{-1/2}$ and $ L_{-2}^{(b^{\prime})} = \frac{1}{2} b_{-3/2}^{\prime} \cdot 
    b_{-1/2}^{\prime}$ the norms of the second and third terms are
    $\frac{1}{4} (\delta_{\mu \mu} \delta_{j j}) = 6$ and $\frac{1}{4}
    (\delta_{\mu \mu} \delta_{k k}) = 5$ respectively. The norm of the
    state given in equation (56) is $-11 + 6 + 5 = 0$

Since $L_{1} = G_{1/2}^{2}$, $L_{1} \mid \Psi \rangle = 0$ implies
$G_{1/2} \mid \Psi \rangle  = 0$. $G_{3/2}$ can be expressed as a commutator of
$L_{1}$ and $G_{1/2}$, so that $G_{3/2} \mid \Psi \rangle =
0$. Further $L_{2} \mid \Psi \rangle = \frac{1}{2} \{ G_{3/2}, G_{1/2} 
\} \mid \Psi \rangle = 0$ and so on, satisfing all the physical state conditions.
\section{Ghosts}
 
For obtaining a zero central charge so that the anomalies
cancel out and natural ghosts are isolated, Faddeev-Popov 
(FP) ghosts \cite{bd7} are introduced. The FP ghost action
is
\begin{equation}
S_{FP}=\frac{1}{\pi}\int (c^+\partial_- b_{++} + c^-
\partial_+b_{--})d^2 \sigma
\end{equation}
where the ghost fields $b$ and $c$ satisfy the anticommutator
relations
\begin{equation}
\{b_{++}(\sigma,\tau), c^+(\sigma^{\prime},\tau)\}
=2 \pi\; \delta(\sigma-\sigma^{\prime})
\end{equation}

\begin{equation}
\{b_{--}(\sigma,\tau), c^-(\sigma^{\prime},\tau)\}
=2 \pi\; \delta(\sigma-\sigma^{\prime})
\end{equation}
and are quantized with the mode expansions
\begin{equation}
c^{\pm}=\sum_{-\infty}^{\infty}c_n\; e^{-in(\tau\pm\sigma)}
\end{equation}

\begin{equation}
b_{\pm \pm}=\sum_{-\infty}^{\infty}b_n\; e^{-in(\tau\pm\sigma)}
\end{equation}
The canonical anticommutator relations for $c_n$'s and
$b_n$'s are
\begin{equation}
\{c_m,b_n\}=\delta_{m+n}
\end{equation}
\begin{equation}
\{c_m,c_n\}=\{b_m,b_n\}=0
\end{equation}

Deriving the energy momentum tensor from the action and making
the mode expansion, the Virasoro generators for the ghosts (G)
are
\begin{equation}
L_m^G=\sum_{n=-\infty}^{\infty}(m-n)\;b_{m+n}\; c_{-n}- a\; \delta_{m}
\end{equation}
where $a$ is the normal ordering constant. These generators 
satisfy the algebra
\begin{equation}
[L_m^G,L_n^G]=(m-n)\;L_{m+n}^G+A^G(m)\; \delta_{m+n}
\end{equation}
The anomaly term is deduced as before and give
\begin{equation}
A^G(m)=\frac{1}{6}(m-13m^3)+2a\;m
\end{equation}
With $a=1$, this anomaly term becomes

\begin{equation}
A^G(m)=-\frac{26}{12}(m^3-m)
\end{equation}
\begin{equation}
B^{G} (r) = - \frac{26}{3} \left ( r^{2} - \frac{1}{4} \right )
\end{equation}

The central charge is $-26$ and cancels the normal ordder $A(m)$ and
$B(r)$ of the $L$ and $G$ generators. Noting that
\begin{equation}
[L^G_m, c_n]=-(2m+n)c_{n+m}
\end{equation}
it is possible to construct an equation for the generator for the current of the ghost sector,
\begin{equation}
G_r^{gh}= \sum_{p} ( \frac{p}{2} - r) c_{-p} G^{gh}_{p+r}
\end{equation}
so that 
\begin{equation}
[L_m^G,G_r^{gh}]=(m/2-r)G^{gh}_{m+r}
\end{equation}

 From Jacobi identity (65)
\begin{equation}
\{ G_{r}^{gh}, G_{s}^{gh} \} = 2 L_{r+s}^{G} + \delta_{r+s} B^{G}(r)
\end{equation}
It immedicately follows that 
\begin{equation}
G_{r}^{gh^{2}} = L^{G}_{2r}
\end{equation}
Since $L_{2r}^{G}$ is well defined, equation (70) has a nonvanishing
solution for $G_{r}^{gh}$. In practice, the products of even number of 
$G_{r}^{gh}$'s occur in calculations and they can be evaluated in
terms of $L_{2r}^{G}$'s.

The total current generator is 
\begin{equation}
G_r=G_r^M+G^{gh}_r
\end{equation}
thus we have the anomaly free Super Virasoro algebra,

\begin{equation}
[L_m,L_n]=(m-n)L_{m+n}
\end{equation}

\begin{equation}
[L_m,G_r]=(m/2-r)G_{r+m}
\end{equation}

\begin{equation}
[G_r,G_s]=2L_{r+s}
\end{equation}
 Thus from the usual conformal field theory we have 
obtained the algebra of a superconformal 
field 
theory. This is the novelty of the present formulation.
The BRST \cite{bd8} charge operator is
\begin{equation}
Q_{BRST}=\sum_{-\infty}^{\infty}L_{-m}^M\;c_m -\frac{1}{2}
\sum_{-\infty}^{\infty}(m-n) :c_{-m}\; c_{-n}\; b_{m+n} :
-a\; c_0
\end{equation}
and is nilpotent for $a=1$. The physical states are such
that $Q_{BRST}\;|phys\rangle=0$.

\section{Fermionic States}

The above deductions can be repeated for Ramond sector~\cite{ramond}. 
We write the main equations. The mode expansion for the fermions are
\begin{equation}
\psi_{\pm}^{\mu,j}(\sigma ,\tau) = \frac{1}{\sqrt 2}
\sum_{-\infty}^{\infty} d_m^{\mu,j} e^{-im(\tau\pm\sigma)}
\end{equation}
\begin{equation}
\phi_{\pm}^{\mu,j}(\sigma ,\tau) = \frac{1}{\sqrt 2}
\sum_{-\infty}^{\infty} d_m^{'\mu,j} e^{-im(\tau\pm\sigma)}
\end{equation}

The generators of the Virasoro operators are
\begin{equation}
L_m^M = L_m^{(\alpha)} + L_m^{(d)} + L_m^{(d')}
\end{equation}
\begin{equation}
L_m^{(d)} = \frac{1}{2}\sum_{n=-\infty}^{\infty}(n+\frac{1}{2} m)
: d_{-n}\cdot d_{m+n} :
\end{equation}
\begin{equation}
L_m^{(d')} = \frac{1}{2}\sum_{n=-\infty}^{\infty}(n+\frac{1}{2} m)
: d'_{-n}\cdot d'_{m+n} :
\end{equation}
and the fermionic current generator is
\begin{equation}
F_m^M = \sum_{n=-\infty}^{\infty}\alpha_{-n}\cdot 
(d_{n+m} + i d'_{n+m}) = \sum_{-\infty}^{\infty}\alpha_{-n}\cdot D_{n+m}
\end{equation}
The Ramond sector Virasoro algebra is the 
same as the NS-sector with the replacement of
G's by F's. It is necessary to define $L_o$ suitably to keep the anomaly 
equations
 the same~\cite{bd5}.

In this Ramond sector, a physical state $\mid \Phi \rangle$ should
satisfy
\begin{equation}
F_{n} \mid \Phi \rangle = L_{n} \mid \Phi \rangle = 0 \; \; \; {\rm
  for} \; \; \; n>0
\end{equation}

The normal order anomaly constant in the anticommutables of the Ramond 
current generators has to be evaluated with care, beacuse the
defination of $F_{0}$ does not have a normal ordering ambiguity. So
$F_{0}^{2} = L_{0}$. Using commutation relation (43) with $G$
replaced by $F$ and the Jacobi Identity we get
\[ \{ F_{r}, F_{-r} \} = \frac{2}{r} \{ [ L_{r}, F_{0} ], F_{-r}
\} = 2 L_{0} + \frac{4}{r} A(r) \]
So
\begin{equation}
B(r) = \frac{4}{r} A(r)
\end{equation}
\begin{equation}
B(r) = \frac{C}{3} (r^{2} - 1), \; \; \; \; r \neq 0
\end{equation}
A physical state in the fermionic sector satisfies
\begin{equation}
( L_{0} - 1 ) \mid \Psi \rangle = 0
\end{equation}
It follows that
\[ (F^{2}_{0} - 1 ) \mid \Psi \rangle  = (F_{0} - 1) (F_{0} + 1) \mid \Psi
\rangle = 0 \]
The Ramond fermonic vacuum is also tachyonic and  could have been  the
supersymmetric partner of the bosonic N - S, tachyonic vacuum. They
are eliminated from Fock spaces by space time supersymmetry.

The construction of `null' physical states becomes much simpler
beacuse all $F_{-m}$ terms can be assigned to $L_{-m}$ terms by the
commutation ruation $F_{-m} = 2 [ F_{0}, L_{-m} ]/m$ and $F_{0}$ has
eigen values which are roots of eigen values of $L_{0}$ acting on the
generic states or states constructed out of the generic states. Thus 
the zero mass null physical state with $L_{0} \mid \tilde{\chi}
\rangle = F_{0}^{2} \mid \tilde{\chi} \rangle = 0$ is simply
\begin{equation}
\mid \Psi \rangle = L_{-1} \mid \tilde{\chi} \rangle
\end{equation}
with $L_{1} \mid \Psi \rangle = F_{1} \mid \Psi \rangle = 0$. The next
 excited state with $(L_{0} + 1) \mid \tilde{\chi} \rangle $ becomes the same as in the bosonic
  sector. Obtained from the condition $L_{1} \mid \Psi \rangle = 0$,
\[ \mid \Psi \rangle = (L_{-2} + \frac{3}{2} L_{-1}^{2}) \mid
\tilde{\chi} \rangle \]
The norm $\langle \Psi  \mid \Psi \rangle = (C - 26)/2$ and vanishes
for $C=26$. It is easy to check that all physical state conditions are 
satisfied. $F_{1} \mid \Psi \rangle = 2$ $[ L_{1}, F_{0} ] \mid \Psi \rangle = 0$
since $L_{1} \mid \Psi \rangle = 0$ and $F_{0} \mid \Psi \rangle =
\mid \Psi \rangle $, $L_{2} \mid \Psi \rangle = F_{1} F_{1} \mid \Psi
\rangle = 0$ and $F_{2} \mid \Psi \rangle = [ L_{2}, F_{0} ] \mid \Psi 
\rangle = 0$. For $C=26$, there are no negative norm states in the
Ramond sector as well.

The ghose curreent in the Ramond sector satisfies the equation 
\begin{equation}
F_{m}^{gh} = \sum_{p} ( \frac{p}{2} -m ) c_{-p} F_{m+p}^{gh}
\end{equation}
so that
\begin{equation}
\left [ L_{m}^{G}, F_{n}^{gh} \right ] = \left ( \frac{m}{2} -n \right 
) F_{m+n}^{gh}
\end{equation}
we can construct $F_{0}^{gh}$ with the help of an anti commuting
object $\Gamma_{n}$ which satisfy 
\begin{equation}
\{ \Gamma_{n}, \Gamma_{m} \} = 2 \delta_{m,n}
\end{equation}
It is important to write $L_{0}^{G}$ in terms of positive integrals as 
\begin{equation}
L_{0}^{G} = \sum_{n=1}^{\infty} n (b_{-n} c_{n} + c_{-n} b_{n} )
\end{equation}
It is found that
\begin{equation}
F_{0}^{gh} = \sum_{n=1}^{\infty} \sqrt{n} \Gamma_{n} ( b_{-n} c_{n} +
c_{-n} b_{n} )
\end{equation}
All other F's can be constructed by the use of the equations of super
Virasoro algebra.

From equation (67) and (86), the ghost current anomaly constant is
$B^{G} (r) = - \frac{26}{3} (r^{2} - 1)$ and cancels out the $B(r)$ of
equation(87). The total current anomalies in both the sectors
vanish.

\section{The Mass Spectrum}

The ghosts are not coupled to the physical states.
Therefore the latter must be of the form (up to null state)\cite{bd9}.
\begin{equation}
|\{n\}\; p\rangle_M \otimes\; c_1|0\rangle_G\label{eq64}
\end{equation}
$|\{n\}\; p\rangle_M$ denotes the occupation numbers and momentum of 
the physical matter states. The operator $c_1$ lowers the
energy of the state by one unit and is necessary for BRST
invariance. The ghost excitation is responsible for lowering
the ground state energy which produces the tachyon. 
\begin{equation}
(L_0^M-1)\;|phys\rangle=0
\end{equation}
Therefore, the mass shell condition is
\begin{equation}
\alpha^{\prime} M^2 = N^B+N^F-1
\end{equation}
where 
\begin{equation}
N^B=\sum_{m=1}^{\infty}\alpha_{-m}\; \alpha_m
\end{equation}
or
\begin{equation}
N^F=\sum_{r=1/2}^{\infty}r\;(b_{-r}\;\cdot  
b_r+b'_{-r}\;\cdot b'_r)\,\,\,\,\, (NS).
\end{equation}

Due to the presence of Ramond and Neveu-Schwartz sectors with periodic and 
anti-periodic 
boundary conditions, we can effect a 
GSO projection ~\cite{bd10} on the mass 
spectrum on the NSR model~\cite{bd13}. Here the
projection should refer to the unprimed and the primed quantas separately. 
The desired projection is
\begin{equation}
G = \frac{1}{4} (1 + (-1)^F)(1 + (-1)^{F'})
\end{equation}
where $ F = \sum b_{-r}\cdot b_r$;    $F' =\sum b'_{-r}\cdot b'_r$ .
This will eliminate the half integral values 
from the mass spectrum by choosing G=1.

For closed strings we have a similar separation 
as in Eq.\ (\ref{eq64}), namely
 the left-handed states will be in the form 
\begin{equation}
|\{ \tilde{n} \}\; p\rangle_M \otimes\; \tilde{c}_1|0\rangle_G
\end{equation}
The mass spectrum can be written as 
\begin{equation}
\frac{1}{2}\alpha ' M^2 = N + \tilde{N} -1 -\tilde{1}
\end{equation}

\section{Modular Invariance}

The GSO projection is necessary for the modular invariance of the
theory. We follow the notation of Seiberg and Witten \cite{bd98}.
Following Kaku \cite{bd13}, the spin structure $\chi (--,\tau)$
for a single fermion is given by
\begin{equation}
\chi(--,\tau) =q^{-1/24} \;Tr\; q^{2\sum_n  n\;\psi_{-n} \psi_n}
=q^{-1/24} \prod_{n=1}^\infty (1+ q^{2n-1}) = 
\sqrt{\frac{\Theta_3(\tau)}{\eta(\tau)}}\;,
\end{equation}
where $\Theta$'s will be the Jacobi Theta functions $\Theta (\theta, \tau)$
\cite{bd98}, $q=e^{i\pi\tau}$ and 
$ \eta (\tau) (2\pi) = \Theta_1^{\prime 1/3}(\tau)$. 
The path integral functions
of Seiberg and Witten for the twenty four unprimed oscillators are
\begin{equation}
A((--),\tau) = (\Theta_3 (\tau)/ \eta(\tau))^{12}\;,
\end{equation}
This is normalised to one. similarly
\begin{equation}
A((+-),\tau) = A((--),\frac{\tau}{1+\tau}) = 
-(\Theta_2 (\tau)/ \eta(\tau))^{12}\;,
\end{equation}
and

\begin{equation}
A((-+),\tau) = A(+-, -\frac{1}{\tau}) =- (\Theta_4(\tau)/\eta(\tau))^{12}\;,
\end{equation}
\begin{equation}
A((++).\tau) =0\;.
\end{equation}
It is easily checked that the sum
\begin{equation}
A(\tau) = (\Theta_3 (\tau)/\eta(\tau))^{12} - 
(\Theta_2 (\tau)/\eta(\tau))^{12}-(\Theta_4(\tau)/\eta(\tau))^{12}
\end{equation}
is modular invariant, using the properties of the theta functions given in
\cite{bd98}.

For the twenty primed oscillators it is not so straightforward because
of the ambiguity of fractional powers of unity. If we prescribe a 
normalization $1=1^{1/2}= \sqrt{e^{2i\pi}}$, then
\begin{equation}
A^\prime ((--),\tau) = (\Theta(\tau)/\eta(\tau))^{10}\;,
\end{equation}
\begin{equation}
A^\prime((+-),\tau) = \sqrt{e^{i\pi}}(\Theta_2(\tau)/\eta(\tau))^{10}\;,
\end{equation}
\begin{equation}
A^\prime((-+),\tau)=\sqrt{e^{i\pi}} (\Theta_4(\tau)/\eta(\tau))^{10}\;,
\end{equation}
\begin{equation}
A^\prime((++),\tau) =0\;.
\end{equation}
The sum $(\Theta_3^{10} (\tau) +\sqrt{e^{i\pi}}\Theta_2^{10} (\tau)
+\sqrt{e^{i\pi}} \Theta_4^{10}(\tau))/\eta^{10}(\tau)$ is also 
modular invariant upto the factor of cube root and fractional roots
of unity. The sum of the modulii is, of course, modular invariant.
It is easy to construct the modular invariant partition function for the
two physical bosons, namely
\begin{equation}
{\cal P}_B(\tau) =(Im \;\tau)^{-2}\Delta^{-2}(\tau){\bar \Delta}^{-2}(\tau)\;,
\end{equation}
in four dimensions \cite{bd96}. For the  normal ordering constant  $-$1 in
$q^{2(L_0-1)}$ , $-$2/12 comes from the bosons 
and $-$44/24 comes from the
fermions adding to  $-$1 for both the NS and the R sectors.

\section{Space-time Supersymmetry}
So far the  drawback of the model, is the existence of the
tachyonic vacuum in both the bosonic and the fermionic sectors. One should 
examine further restictions imposed on the Fock space due to the
 space time supersymmetric algebra. It is already been noted in reference
\cite{bd22}, that a standard like model $SU(3) \times SU(2) \times
U(1) \times U(1)$ can be space time supersymmetric. The supersymmetric
charge $Q$ should be such that
\begin{equation}
\delta X^{\mu} = \left [ X^\mu , \bar{Q} \cdot \epsilon \right ]
= \bar{\epsilon} \cdot \psi^{\mu}
\end{equation}
and
\begin{equation}
\delta \psi^{\mu} = \left [ \psi^{\mu}, \bar{Q} \cdot \epsilon \right
] = -i \rho^{\alpha} \partial_{\alpha} X^{\mu} \epsilon
\end{equation}
A simple inspection shows that
\begin{equation}
\bar{Q} = -\frac{i}{\pi} \int_{0}^{\pi} d \sigma \psi_{\mu} \rho^{\alpha} \rho^0
\partial_{\alpha} X^{\mu}
\end{equation}
leading to 
\begin{equation}
Q^{\dagger} = -\frac{i}{\pi} \int_{0}^{\pi} d \sigma \psi_{\mu} \rho^{\alpha} \rho^0
\partial_{\alpha} X^{\mu}
\end{equation}
and
\begin{equation}
Q = \frac{i}{\pi} \int_{0}^{\pi} \rho^0 \rho^{\alpha}{}^{\dagger} \partial_{\alpha}
X^{\mu} \psi_{\mu} d \sigma
\end{equation}
By a somewhat lengthy calculation it is decuced that 
\begin{equation}
\sum_{\alpha} \{ Q_{\alpha}^{\dagger}, Q_{\alpha} \} = 2H
\end{equation}
where $H$ is the Hamiltonian of the system. It follows that for the
ground state $\mid \Phi_{0} \rangle$ in the Fock space 
\begin{equation}
\sum_{\alpha} \mid Q_{\alpha} \mid \phi_0 \rangle \mid^2 = 2 \langle
\phi_0 \mid H \mid \phi_0 \rangle \geq 0
\end{equation}
It is essential that the tachyonic vacuum should be discarded from
the physical Fock space and relegated
to the ghost space to satisfy this result of the  exact space-time
supersymmetry. This has to be done in addition to the GSO projection. The
ground state is massless. 

Admissible $SO(6) \times SO(5)$ symmetric Fock space states are

\[ NS~{\rm eigenstates:} \: \: \: \prod_{n, \mu} \prod_{m, \nu} \{
\alpha_{-n}^{\mu} \} \{ B_{-m}^{\nu} \} \mid 0 \rangle \]
\[  R~{\rm eigenstates:} \: \: \: \prod_{n, \mu} \prod_{m, 0} \{
\alpha_{-n}^{\mu} \} \{ D_{-m}^{\nu} \} \mid 0 \rangle u \]

Both the  tachyons must be omitted and GSO projection is implied for the N S eigenstates.Let us construct the zero mass modes. The tachyonic vacuum will be
denoted by $|0\rangle$ and the zero mass ground state by $\mid \phi_0
\rangle$. We start with the supergravity multiplet. The ground state
\begin{equation}
B^{\mu}_{-1/2} B^{\nu}_{-1/2} \mid 0 \rangle \epsilon_{\mu \nu}
\end{equation}
has zero mass. Due to the physical state conditions $G_{1/2} \mid
\phi_{0} \rangle = 0$
\begin{equation}
p^{\mu} \epsilon_{\mu \nu} = p^{\nu} \epsilon_{\mu \nu} = 0 
\end{equation}
It describes a massless antisymmetric tensor $A_{\mu \nu} = 1/2
(\epsilon_{\mu \nu} - \epsilon_{\nu \mu})$, which turns out to be a
pseudoscalar, a massless scalar $\epsilon_{\mu \mu}$ of spin $0$ and a 
massless symmetric terms  of spin 2: $\epsilon_{\mu \nu} - 1/2
(\epsilon_{\mu \nu} + \epsilon_{\nu \mu})$, which is traceless.

The other zero mass spinonial states are
\begin{equation}
\alpha_{-1}^{\mu} \mid 0 \rangle u_{1 \mu}
\end{equation}
\begin{equation}
D_{-1}^{\mu} \mid 0 \rangle u_{2 \mu}
\end{equation}
$u_{1 \mu}, u_{2 \mu}$ are spinor four vectors and are distinguished
by \cite{bd5},
\begin{equation}
\gamma_{5} u_{1 \mu} = u_{1 \mu}
\end{equation}
\begin{equation}
\gamma_{S} u_{2 \mu} = - u_{2 \mu}
\end{equation}
We shall consider them together as a four component spin vector
$u_{\mu}$. The condition $F_0 \mid \phi_0 \rangle = 0$, $F_1 \mid
\phi_0 \rangle = 0$, $L_1 \mid \phi_0 \rangle = 0$ lead to the
condition
\begin{equation}
\gamma \cdot p u_{\mu} = p^{\mu} u_{\mu} = \gamma^{\mu} u_{\mu} = 0
\end{equation}
This state  contains not only a spin $3/2$ but also a spin $1/2$
state. They can be projected out. The details have been given by GSO
in reference \cite{bd10}.

We now count the number of physical degrees of freedom \cite{bd10}
\begin{tabbing}
fd;klfds;klfds;klfdsk;lfds;klfds;lkfds\= klfdlkjfdskjlfdlkjfdslkjfdlkfdlkjfdklj\= lkfdsjlkfdsklfdskljfdsjlkf\= \kill
Graviton\> 2 degrees of freedom:\> $\rho^{a}_{\mu}$\>\\
Dilaton, $\epsilon_{\mu \mu}$, \> 1\> A\>\\
Antisymmetric tensor\> 1\> B\>\\
Spin 3/2\> 2\> $u_{\mu}$\>\\
spin 1/2\> 2\> u\>
\end{tabbing}

The numbers of the fermions and the bosons are equal. They can be grouped
together as the gravitational $(\rho^{a}_{\mu}, u_{\mu})$ and 
the matter $(A,B,u)$ multiplets.

The massless ground state vector is represented by
\begin{equation}
\alpha^{\mu}_{-1} \mid 0 \rangle \epsilon_{\mu} (p)
\end{equation}
Here, because of the $L_0$ condition, $p^2 \epsilon_\mu = 0$: The
constraint $ L_1 \mid \phi_0
\rangle = 0$ gives the Lovertz condition $p \cdot \epsilon = 0$. The
external photon polarisation vector can be subjected to an on shell
gauge transformation $\epsilon_{\mu}(p) \rightarrow \epsilon_{\mu} (p) 
+ \lambda p_{\mu}$. Therefore the state
\begin{equation}
p_{\mu} \alpha^{\mu}_{-1} \mid 0 \rangle \lambda = L_{-1} \mid 0
\rangle \lambda
\end{equation}
decouples from the physical system. There are only two degrees of
freedom left. However, from the Ramond sector we have the spinor (gaugino)
\begin{equation}
p_{\mu} \alpha^{\mu}_{-1} \mid 0 \rangle u(p) = F_{-1} \mid 0 \rangle u(p)
\end{equation}
with $\gamma .p$ $u(p) = 0$ from the physical state condition. Further, as already noted, $\gamma_{5} u(p) = u(p)$. So the member of
the fermionic degrees of freedom is again two, just like the vector
boson. Thus for all the zero mass states the bosonic and the fermonic
degrees of freedom are equal.

\section{Approach to standard model}

One of the main motivation of constructing this superstring is to show 
that the internal symmetry group makes a direct contact with the
standard model  which explains all available experimental
data with a high degree of accuracy. Since there are forty four
fermions, the internal symmetry group was $SO(44)$. We divided these
fermions in groups of eleven where each group was characterised by a
space-time index $\mu = 0, 1, 2, 3$. All the four groups are similar,
but not idential. The states when acted upon by creation/anihilation
operators with $\mu = 0$ are eliminated due to Virasoro constraints
and the states with  negative norm  are absent. The other three groups of
eleven, $\mu = 1, 2, 3$ are identical and have $SO(11)$
symmetry. These may be construed to be the three generations of
the standard model.

By partitioning further, each $SO(11)$ has been broken up into a
product of $SO(6)$ and $SO(5)$. so the internal suymmetry of the model 
is $SU(4) \times SO(5)$. According to Slansky \cite{bd12}, $SO(5)$ can 
break to $SU(2) \times SU(2)$. Thus we are led to the Pati-Salam group
\cite{23}, $SU(4) \times SU_L (2) \times SU_R (2)$. The most convinent 
scheme of descending to the standard model is
\begin{eqnarray}
SO(11) \longrightarrow && SO(6) \times SO(5) \nonumber \\
&& \downarrow M_x \nonumber \\
&& SU(4) \times SU_L (2) \times SU_R (2) \nonumber \\
&& \downarrow M_R \nonumber \\
&& \times \nonumber \\
&& \downarrow M_S \nonumber \\
&& SU_C (3) \times SU_L (2) \times U_Y (1)
\end{eqnarray}

Such a scheme and similar ones have been extensively studied
\cite{24}. Invoking charge quantisation, $SU(4)$ may be broken to
$U_{B-L} (1) \times SU_C (3)$ and subsequently $U_{B-L} (1)$ may
squeeze with $SU_R (2)$ to yiend $SU_Y (1)$. Unification mass is $M_X
= M_{GUT}$, the left-right symmetry breaks at $M_R$ and
supersymmetry is broken at $M_S = M_{SUSY}$. The renormalisation
equations for the evolution of the coupling constants are easily
written down \cite{25}.

We denote $\alpha_i = g_i^2 / 4  \pi$ where $g_i$ is the constant
related to the $i^{th}$ group, $\alpha_G = g_{u}^{2} / 4 \pi$ where
$g_u$ is the coupling constant at the GUT energy and $t_{XY} =
\frac{1}{2\pi} \log_e M_X / M_Y$. The lowest order evolution equations 
are
\begin{equation}
\alpha_{3}^{-1} (M_Z) = \alpha_{G}^{-1} + b_3 t_{SZ} + b_{3s} t_{RS} +
  b_{4s} t_{XR}, 
\end{equation}
\begin{equation}
\alpha_{2}^{-1} (M_Z) = \alpha_{G}^{-1} + b_2 t_{SZ} + b_{2s} t_{RS} +
  b_{2s} t_{XR},
\end{equation}
and
\begin{equation}
\alpha_{1}^{-1} (M_Z) = \alpha_{G}^{-1} + b_1 t_{SZ} + b_{1s} t_{RS} +
  ( \frac{2}{5} b_{4s} + \frac{3}{5} b_{2s} ) t_{XR}.
\end{equation}
$b_i$ and $b_{is}$ are the well known non susy and susy coefficients
of the $\beta$-function respectively. The experimental values at $M_Z
= 91.18$ GeV are calculated to be \cite{26}
\begin{equation}
\alpha_{1}^{-1} = 59.036, \alpha_{2}^{-1} = 29.656, \alpha_{3}^{-1} =
7.69
\end{equation}
To these, we add the expected string unification value
\begin{equation}
M_X = M_{GUT} = M_{string} = g_U (5 \times 10^{17}) GeV
\end{equation}

We have four unknown quantities to calculate from the four known
values, equations (135) and (136).

Notice that the quantities, $b_1 - 3/5 b_2 = 6, b_{1s} - 3/5 b_{2s} = 6,
b_3 = -7, b_{3s} = -3$ and $b_{4s} = -6$ are independent of the
required number of Higgs doublets. so we rewrite the above three
equations as
\begin{equation}
\alpha_{1}^{-1} - 3/5 \alpha_{2}^{-1} - 2/5 \alpha_{3}^{-1} = 8.8
t_{SZ} + 7.2 t_{RS}
\end{equation}
\begin{equation}
\alpha_{1}^{-1} - 3/5 \alpha_{2}^{-1} = 2/5 \alpha_{G}^{-1} + 6 t_{SZ} 
+ 6 t_{RS} - 2.4 t_{XR}
\end{equation}
\begin{equation}
\alpha_{3}^{-1} = \alpha_{G}^{-1} - 7 t_{SZ} - 3 t_{RS} - 6 t_{XR}
\end{equation}
The solutions are $M_{SUSY} = 5 \times 10^9 $Gev, $M_R = 5 \times
10^{14}$Gev, $M_X = 2.87 \times 10^{17}$ and $g_u = 0.566$. With the
value of $M_R$ found here, the mass of the left-handed tau neutrino
\cite{28} is calculated to be about $1/25 ev$. following references
\cite{24} and \cite{27}. We have used $m_{top} (M_R) \cong 140
Gev$ in the formula for the neutrino mass $m_{\nu \tau}$,
\begin{equation}
m_{\nu \tau} = - \frac{m_{\rm top}^{2}}{ M_R}
\end{equation} 
This is a very important result of the model.
\section{Conclusion}
It is remarkable that we have been able to discuss physics from the Planck
scale to the Kamiokanda neutrino scale within the same framework. The
starting point has been a Nambu-Gatto string in four dimensions to
which forty four Majorana neutrinoes in groups of four have been
added. The resulting string has an action which is
supersymmetric. Super-Virasoro algebra for the energymomentum tensor and 
current generators is established. Conformal ghosts are introduced
whose contributions cancell the anomalies. BRST charge is explicitly 
constructed.

The main drawback of the theory is the presence of the two tachyons in 
the bosonic and fermionic sector even after GSO projections. Since the 
space-time supersymmetry algebra is satisfied by the action, the two
tachyons must be discarded from the physical spectrum.

The internal symmetry of the string is $SO(6) \times SO(5)$ which
breaks to the Pati-Salam group $SU(4) \times SU_L (2) \times SU_R (2)$ 
at the string scale. The left right
symmetry and supersymmetry are broken at intermediate mass scales. By the usual see-saw mechanism, the left handed 
neutrino develops a small mass of  about $\frac{1}{25}$ ev. Finally
the descent is complete at $SU_C (3) \times SU_L (2) \times U_Y
(1)$. There is no gap left between $M_{GUT}$ and $M_{string}$ by
choice.\\

\centerline{{\underline Acknowledgement}}

I have profited from discussions with Dr. J.  Maharana and Dr. S. Mahapatra.
I thank Sri D. Pradhan for computer compilation and the Institute of
Physics for providing Library and  Computer  facilities.


\end{document}